\documentstyle[12pt,worldsci]{article}
 1
\def\MPL #1 #2 #3 {Mod.~Phys.~Lett.~{\bf#1},\ #2 (#3)}
\def\NPB #1 #2 #3 {Nucl.~Phys.~{\bf#1},\ #2 (#3)}
\def\PLB #1 #2 #3 {Phys.~Lett.~{\bf#1},\ #2 (#3)}
\def\PR #1 #2 #3 {Phys.~Rep.~{\bf#1},\ #2 (#3)}
\def\PRD #1 #2 #3 {Phys.~Rev.~{\bf#1},\ #2 (#3)}
\def\PRL #1 #2 #3 {Phys.~Rev.~Lett.~{\bf#1},\ #2 (#3)}
\def\RMP #1 #2 #3 {Rev.~Mod.~Phys.~{\bf#1},\ #2 (#3)}
\def\ZP #1 #2 #3 {Z.~Phys.~{\bf#1},\ #2 (#3)}
\def\IJMP #1 #2 #3 {Int.~J.~Mod.~Phys.~{\bf#1},\ #2 (#3)}
\begin{document}
%
%
\rightline{\vbox{\halign{&#\hfil\cr
&ANL-HEP-CP-95-28\cr
&PSU/TH/160\cr
&June 1995\cr
}}}
%
%
\begin{center}{{\bf PROSPECTS FOR
SPIN PHYSICS AT RHIC\footnote{To appear in the
proceedings of
{\it The International Symposium on Particle
Theory and Phenomenology,
Iowa State University, May 22-24, 1995.}}
\\}
\vglue 0.6cm
{R. W. Robinett\footnote{Work supported in part
by the Department of
Energy, Contract W-31-109-ENG-38.} \\}
\baselineskip=13pt
{\it High Energy Physics Division, Argonne
National Lab,
Argonne IL 60439 USA
\\}
\vglue 0.1cm
and \\
{\it Department of Physics,
Penn State University, University Park PA 16802
USA\footnote{Permanent address.}
\\}
\vglue 0.2cm
ABSTRACT}
\end{center}
{\rightskip=3pc
 \leftskip=3pc
 \baselineskip=12pt
 \noindent
The proposal to perform polarized proton-proton
collisions at
collider energies at RHIC is reviewed.
After a brief reminder
of the desirability of high energy
spin physics measurements,
we discuss the machine parameters
and detector features
 which are taken
to define a program of spin physics at RHIC.
Some of the many physics
processes which can provide information
on polarized parton
distributions and the spin-dependence
of QCD and the electroweak
model at RHIC energies are discussed.
\vglue 0.6cm}

\section{Motivation for collider energy spin-physics}

The parton structure of the proton, first revealed by  deep
inelastic scattering (DIS), continues to
excite interest as  experiments \cite{mallik} extending that
program to lower values of $x$ now
probe the QCD structure of the proton in new regimes.
In a similar way, polarized DIS experiments have
shown that the spin structure of the proton
is more complex than
envisaged by naive considerations based
on the constituent quark
model.  The EMC \cite{emc} measurements
of $g_1^{(p)}(x)$ first suggested
that the total contribution of the
light quarks to the proton
spin was surprisingly small ($\Delta \Sigma =
\Delta u + \Delta d + \Delta s <<1$) and
that there is a
non-negligible contribution from
sea quarks ($\Delta s \neq 0$).
Motivated by these results, new experimental
tests designed to
accurately measure
the spin-dependent structure functions of
both the neutron
and proton (thereby testing the Bjorken sum rule)
were undertaken.
The results of the SMC \cite{smc} and
SLAC \cite{e142e143}
experiments have been consistently
analyzed \cite{ellis-karliner} and
imply that:
\begin{itemize}
\item The Bjorken sum rule has been
verified to $\sim\!10\%$ or,
equivalently, the value of $\alpha_S(Q^2\!=\!M_Z^2)$
extracted from the sum rule
(including radiative corrections)
is consistent
with values extracted from many other QCD processes.
\item The individual Ellis-Jaffe sum rules
for the proton and neutron
are best fit with a non-zero value
of $\Delta s$, indicating a
non-zero sea quark polarization.
\item The light quark contribution to
the total proton spin,
$\Delta \Sigma = \sum_q \Delta q$, which is constrained
by the angular momentum sum rule,
\begin{equation}
\frac{1}{2} = \frac{1}{2}\Delta \Sigma + \Delta G +
\langle L_z \rangle
\label{angular-momentum-sum-rule}
\end{equation}
is roughly $\Delta \Sigma = 0.31\pm 0.07$, indicating
that a large
fraction of the proton spin arises from
polarized gluons or
orbital angular momentum.
\end{itemize}

The lack of flavor separation in polarized
DIS\footnote{Recent measurements of semi-inclusive
and inclusive spin asymmetries by the SMC
collaboration\cite{SMC-flavor-dis} have, for
the first time, allowed for a more direct separation
of the valence $u$ and $d$ and non-strange sea
quark spin fractions as a function of $x$.}
and the looser constraint on $\Delta G$ imposed by
Eqn.~\ref{angular-momentum-sum-rule} as
compared with the
corresponding sum rule for linear momentum
immediately imply
that more experimental input to better
determine $\Delta G(x)$ and
$\Delta \overline{q}(x)$ would be
very useful.  Many of the most
important new constraints on
unpolarized parton distributions
\cite{tung} now come from hadron collisions,
so it is natural to
study the extent to which a program of
high-energy
polarized hadron collisions
can give information on the spin-dependent
parton distributions.
Such a program would also provide an opportunity
for the
 systematic study of the spin-dependence of QCD
and the electroweak theory
for the first time.

Any program of collider-energy spin physics would
benefit from the following aspects:
\newcounter{temp}
\begin{list}
{(\roman{temp})}{\usecounter{temp}}
\item  High enough energy to ensure that a
leading-twist,
perturbative QCD description is unambiguously applicable.
Previous polarized $pp$ collisions at lower-energy,
fixed target facilities
have shown dramatic spin effects, but are not obviously
reliably describable by perturbative QCD,
\item High luminosity is important for any
program attempting to
measure possibly small asymmetries in
cross-sections due to spin-effects,
\item Large polarization in both beam and target
(so that the fundamental
partonic-level hard-scattering spin dependence
is not diluted) is
valuable.  While collisions in which only one
hadron is polarized
can be used to probe spin-dependence when parity
violation is
present (as in $W/Z$ production), doubly
polarized collisions
can make maximal use of the intrinsically large
partonic level spin-spin asymmetries
($\hat{a}_{LL}$) in QCD to
provide information on the polarized parton
distributions.
In addition, the possibility of obtaining
both longitudinal
polarization (which is the type probed by
polarized DIS) {\it and}
transverse spin (the effects of which decouple in DIS)
is highly desirable,
\item If a multi-purpose collider-type
detector is not available,
the detectors should be versatile enough
to still allow for
a comprehensive program of spin physics
measurements,
\item Finally, one would like such a
program to fit naturally
and economically into the existing plans
and likely funding profiles for
high energy physics research.
\end{list}
\section{RHIC spin program defined}

Building on earlier studies of polarized
proton-proton collisions
for the then proposed ISABELLE collider
\cite{courant} and
motivated by the first successful tests
of the Siberian snake
concept \cite{krisch} (the technology required
to maintain
proton polarization in circular
accelerators), the following
parameters \cite{workshop,particleworld}
for a program of polarized $pp$
collisions at RHIC
are now thought to be achievable:
\begin{itemize}
\item  Variable center-of-mass energy
in the range
$\sqrt{s} = 50-500\;GeV$.
\item  Luminosities up to ${\cal L} =
2\times 10^{32}\;
cm^{-2}\sec^{-1}\;
(8\times 10^{31}\;cm^{-2}\sec^{-1} )$  at
$\sqrt{s} = 500\;GeV\;(50\;GeV)$.
\item   Both longitudinal and transverse
polarization available at
the intersection regions of both large detectors with
$P^2 \approx (0.7)^2 = 0.5$.
\item   Rapid switching of the proton
polarization can effectively eliminate
systematic errors.
\item  Roughly $10$ weeks running time
for an integrated luminosity
of $800 \, (320)\;pb^{-1}$ at $\sqrt{s} =
500\;GeV \, (50\;GeV)$
per year.
\end{itemize}

The two large heavy-ion detectors approved
for RHIC are STAR
(which is a large-acceptance, TPC-based
tracking detector)
and PHENIX (which emphasizes
$\gamma$, $e$, and $\mu$ detection).
Both detector groups, joined by the RHIC
Spin Collaboration (RSC),
have put forward a comprehensive
proposal \cite{proposal}
to perform a program of
collider energy spin-physics measurements
which was
approved in 1993 as RHIC experiment R5.
The STAR
detector would be enhanced by an upgrade
to include a barrel
EMC to aid in jet and direct photon
detection.  The RIKEN/SPIN
group from Japan has promised $20\,M\$$ to
provide for the
magnets required for obtaining spin
(i.e., the Siberian snakes
and spin rotators in the RHIC tunnel)
as well as for an upgrade
\cite{phenix-muons}
of the muon tracking capability of the
PHENIX detector which
would dramatically enhance it's ability
to see Drell-Yan pairs
and weak bosons as well as heavy quarks.
This additional source of funding, along
with the recent
successful testing\cite{huang} of the
partial snake technology  in the AGS ring
required for injection of polarized
protons into RHIC,  are, perhaps, the two
most important developments
in the last year for the continued
development of the RHIC spin program.
\section{Spin-dependent collider processes}
The full use of a hadron collider in which
both beams are longitudinally
polarized comes from measurements of the
spin-spin asymmetry, defined via
\begin{equation}
A_{LL} = P_1 P_2
\left(\frac{\Delta \sigma}{\sigma} \right)
= P_1 P_2 \left(\frac{\sigma(++) - \sigma(+-)}
{\sigma(++) + \sigma(+-)}
\right)
\label{observable-asymmetry}
\end{equation}
where $P_{1,2}$ are the beam polarizations
and $\sigma(+,\pm)$ indicates
the differential cross-section for any
observable quantity for the case
where the first beam is polarized in the $+$
direction while the second
is in either polarization state.  The spin-dependent
cross-sections are
given by
\begin{equation}
\Delta \sigma \;\; \propto \;\; \sigma(++) - \sigma(+-)
= \sum_{i,j}\; \int \int \,
\Delta f_{i}(x_1)\;\Delta f_{j}(x_2)\;
\hat{a}_{LL}^{(i,j)}\;d\hat{ \sigma}_{i,j}
\end{equation}
where $\Delta f_{i,j}(x)$ are the
polarized parton densities
and the partonic-level spin-spin asymmetry is defined by
\begin{equation}
\hat{a}_{LL}^{(i,j)} =
\frac{d\hat{\sigma}_{i,j}(++) \!-\! d\hat{\sigma}_{i,j}(+-)}
{d\hat{\sigma}_{i,j}(++) \!+\! d\hat{\sigma}_{i,j}(+-)}
\end{equation}
for each contributing subprocess.
Many of the $\hat{a}_{LL}$ for subprocesses
contributing to familiar
QCD processes (such as direct $\gamma$, jet,
and Drell-Yan production
are as large as possible, close to $|\hat{a}_{LL}|
\approx 0.8\!-\!1$)
Experimental measurements of $A_{LL}$ thus
give information on the spin-dependent parton
distributions as well
as the intrinsic spin-dependence.  Single spin
asymmetries,  defined as
\begin{equation}
A_{L} = P_1 \left(
\frac{\sigma(+) \!-\! \sigma(-)}
{\sigma(+) \!+\! \sigma(-)}\right)
\end{equation}
are also possible if there is intrinsic parity
violation in the process,
as in weak boson production.  The intrinsic
spin-dependence and
sensitivity to polarized parton distributions
for many standard model
processes have been studied in the context of
the RHIC spin program
and we give some examples below.

\subsection{Direct $\gamma$ production}
Given the important role that direct photon
production has played in
the determination of the unpolarized gluon
distribution, it is not
surprising that it is touted \cite{berger-qiu}
as an effective tool for probing
$\Delta G(x)$.  At leading order, the
Compton diagram $qg \rightarrow
q \gamma$ dominates (especially in $pp$ collisions)
and measurements
of the asymmetry in the $\gamma$ plus away-side-jet
cross-section
provide an almost direct measure of $\Delta G(x)/G(x)$.
Even without
such identification, strong constraints on the
magnitude of $\Delta G(x)$
can be obtained. The next-to-leading-order (NLO)
corrections have
been performed by two groups \cite{nlo-direct-gamma}
who find
that the LO predictions for the
spin-spin asymmetry
are perturbatively stable.

\subsection{Jet production}
The number of subprocesses contributing to
jet production \cite{jets} is larger,
 with the relative importance of individual QCD
subprocesses ($gg$, $qg$, and $qq$ initial states)
varying with
$p_T$ or $M_{jj}$.  The observation of
$A_{LL}$ at large
$p_T$, where the $qq$ subprocesses
dominate and where the polarization
of the valence quarks is known to be
large, should provide a benchmark
test of the QCD predictions for
 collider spin physics.
The partonic-level asymmetries for all of the
leading-order QCD processes are
 large \cite{soffer-review}
as well as those for the $2 \rightarrow 3$
processes \cite{rick-3-jet} which
could contribute at NLO.  While the NLO
corrections to jet
production in polarized hadron collisions
have not yet been completed,
the necessary NLO helicity-dependent
$2 \rightarrow 2$
matrix-elements have been
calculated\cite{bern}.

\subsection{Heavy quark production}

Armed with the NLO corrections to
heavy quark production, it has
been suggested \cite{berger-meng-tung}
that $b$-quark production can be used
to provide
constraints on the unpolarized gluon
densities in the proton.
Early discussions of heavy quark production
in polarized
$pp$ collisions, which focused on spin
asymmetries in the total
cross-section\cite{contogouris} , have
been extended
\cite{rick-karliner} to include the $p_T$
dependence which was found
to be dramatic, leading to a maximally large value of
$\hat{a}_{LL} = -1$ for $p_T >> M_Q$.
 It has
been suggested\cite{rick-karliner} ,
however, that NLO corrections
could change the LO picture.  Recent NLO
calculations of
$Q\overline{Q}$ production in polarized
$\gamma \gamma$ collisions
\cite{nlo-b}
tend to support this conjecture
 where the dominant $(+-)$ helicity combination
is perturbatively stable while the highly
suppressed (at leading order)
$(++)$ contribution receives a very large
 NLO correction.  The
enhanced PHENIX capability for $b$-quark
 detection makes further study
of this process highly desirable.
It could well provide an interesting
test of the subtle interplay of LO
and NLO spin-dependence in QCD.

\subsection{Other processes probing $\Delta G(x)$}
There have been a number of studies of
 other processes which are sensitive
to $\Delta G(x)$, namely $3$-jet  \cite{rick-3-jet}
and $4$-jet \cite{rick-4-jet} production, $\psi$
production
at both low \cite{cortes,contogouris,rick-low-pt} and
high \cite{rick-high-pt} $p_T$,
double-photon production\cite{gamma-gamma} ,
and $\psi +\! \!\gamma$
production \cite{dkim}.

\subsection{Probing $\Delta \overline{q}(x)$}

The Drell-Yan process is the standard hadronic
 probe of the
anti-quark distribution and many studies of its
 role in polarized
$pp$ collisions, both in the context of $\gamma^*$
\cite{drell-yan,nlo-drell-yan}
 and $W/Z$ \cite{w-z-rhic-soffer,zphys}
production have appeared.  Such studies are timely as RHIC
may be the best laboratory for the
 study of the $SU(2)$
structure of the polarized and unpolarized
\cite{halzen} sea quark distributions.

\subsection{Measuring transversity distributions}
The other leading twist-2 observable
 corresponding to the
distribution of transverse spin
 in a proton \cite{jaffe-ji}
decouples from DIS because of a
 helicity mismatch.  No
experimental information is currently
 available on its magnitude,
although bag model studies suggest that it
 is of the same
order as the corresponding longitudinal
 quantity.  The
`transversity' distribution gives a
leading-twist
transverse spin asymmetry ($A_{TT}$)
 in polarized Drell-Yan production,
but in $pp$ collisions it
probes a combination of unknown
 quark {\it and} anti-quark
`transversities'; further information to
 help separate the
two may  come from transversely
polarized $Z$ production\cite{w-z-rhic-soffer}.
Jet production at large $p_T$, where $qq$
 scattering dominates, can
show a non-vanishing $A_{TT}$ depending
 on the valence quarks alone;
the corresponding partonic-level spin-spin
 asymmetries, $\hat{a}_{TT}$,
are, however,  smaller than the corresponding
$\hat{a}_{LL}$ for both 2-jet \cite{sivers}
 and $3$-jet
\cite{rwr4} production due to a color
 mismatch in the required
interference diagrams.  High statistics
 experiments will likely be
needed \cite{artru} to study the signal
 in the region where the transversity
distribution is important in this sector.

\section{Summary}  A polarized
proton-proton facility at RHIC
will be a unique laboratory for the
measurement of the
longitudinal and transverse spin-dependent
parton distributions and
for testing the spin-dependence of QCD and the
 electroweak interactions.
It is a more versatile program for measuring
$\Delta G(x)$ than other
more specialized proposals and is nicely
complementary to the RHIC
program of heavy ion physics; hadron
 spin-dependence and structure
function physics are now sometimes
 considered as one of the `new'
directions in nuclear physics and
share important non-perturbative
physics aspects with the deconfinement
 transition expected as the
ultimate goal in
 the quest for the quark-gluon plasma.
 The addition of a spin physics
program to the RHIC project
provides a large physics benefit at a small incremental
cost.
\bibliographystyle{unsrt}

\end{document}